\documentclass{article}

\usepackage{arxiv}

\usepackage[utf8]{inputenc} 
\usepackage[T1]{fontenc}    
\usepackage{hyperref}       
\usepackage{url}            
\usepackage{booktabs}       
\usepackage{amsfonts}       
\usepackage{nicefrac}       
\usepackage{microtype}      

\usepackage{amssymb}
\usepackage{amsmath}
\usepackage[figuresright]{rotating}
\usepackage{caption}
\usepackage{subcaption}
\usepackage{multirow}
\usepackage[ruled]{algorithm2e}

\usepackage{xcolor}
\definecolor{RoyalBlue}{RGB}{65, 105, 225}

\title{Mutation-Based Adversarial Attacks on \\Neural Text Detectors}

\author{Gongbo~Liang$^*$,~~Jesus~Guerrero,~~Izzat~Alsmadi$^*$\\ 
		Department of Computing and Cyber Security\\ 
        Texas A\&M University -- San Antonio\\ 
        \{gliang, ialsmadi\}@tamusa.edu
		}

\begin{document}
\maketitle

\begin{abstract}
Neural text detectors aim to decide the characteristics that distinguish neural (machine-generated) from human texts. To challenge such detectors, adversarial attacks can alter the statistical characteristics of the generated text making the detection task more and more difficult. Inspired by the advances of mutation analysis in software development and testing, in this paper, we propose character- and word-based mutation operators for generating adversarial samples to attack state-of-the-art natural text detectors. This falls under white-box adversarial attacks. In such attacks, attackers have access to the original text and create mutation instances based on this original text.  The ultimate goal is to confuse machine learning models and classifiers and decrease their prediction accuracy. 
We introduced a general framework for building the character- and word-level mutation operators. Several operators were demonstrated and evaluated using the text captions of the MS COCO2017 dataset and state-of-the-art neural language models. We believe the proposed mutation-based adversarial attacks can be used as a systematic way to evaluate the robustness of any language analysis models.
\end{abstract}


\section{Introduction}
Modern neural network (NN) models can generate textual and imagery content looking as if genuine, created by humans ~\cite{vaswani2017attention,radford2019language,Karras2019style,workman2022revisiting}. Such computer-based, automatic content generation benefits society in many domains, from the medical field~\cite{mihail2019automatic,liu2022spatiotemporal} to business and education~\cite{kaczorowska2019chatbots,kerlyl2006bringing}. However, it also makes it easier to generate human-like content at a large scale for nefarious activities, from spreading misinformation to targeting specific groups (e.g., for political agenda) ~\cite{stiff2022detecting,wolff2020attacking,alsmadi2021adversarial}. One interesting recent example is OpenAI ChatGPT~\cite{chatgpt}; such a large NN language model is expected to be disruptive and significantly impact many domains.
Researchers are actively working on the generation and detection of imagery synthesized or manipulated content~\cite{dang2020detection,zhang2019defense,ciftci2020fakecatcher}. Unfortunately, the textual counterpart is still under-researched~\cite{stiff2022detecting,wolff2020attacking}.

Inspired by the advances of mutation analysis in software testing and the idea of ``the boiling frog syndrome", we propose a general mutation-based framework that generates mutation text for attacking pre-trained neural text detectors. Unlike the popular AI-based text generation methods, such as Transformer~\cite{vaswani2017attention}, BERT~\cite{devlin2018bert}, or GPT-3~\cite{brown2020language}, which generate contents in a less controlled and open-ended environment, the output of our method is precise and close-ended, which enables our mutation-based text generation method to evaluate the robustness of any language analysis model systematically. We demonstrate this evaluation by using the output of our method as adversarial samples to attack the state-of-the-art neural text detector model, the RoBERTa-based~\cite{liu2019roberta} detector released by OpenAI.

To generate text in a precise and closed-ended fashion, an input text sequence and a mutation operator are needed. The text generated by our method is changed slightly from the original input text based on the mutation, for instance, given a text sequence of \texttt{"an apple"} and a mutation operator replacing the English letter \texttt{"a"} in the article \texttt{"an"} with the Greek letter $\alpha$, our method gives the output of \texttt{"$\alpha$n apple"}. Ideally, a robust language analysis model should tolerate small changes like this. However, our experimental results show that such minor changes easily fool the state-of-the-art, RoBERTa-based detector. For some test cases, the performance is dropped to only $0.07$ AUC (the area under the receiver operating characteristic curve).

Through this work, we also demonstrated a random removing (RR) training strategy that can be used to improve the robustness of language analysis models. Our experiments show that the robustness of the RoBERTa-based detector can be improved up to $9.40\%$ when the RR training strategy is applied during the fine-tuning stage.

We believe that mutation-based adversarial attacks offer a systematic way to evaluate the robustness of language analysis models that have not been evaluated before. The random removing training strategy is useful to improve model robustness without involving any additional data or a significant amount of work. To summarize, our paper presents the following intuitive contributions:
\begin{itemize}
    \item Introducing the mutation-based text generation strategy for systematically evaluating the robustness of language analysis models.
    \item Proposing, generating, and evaluating several mutation-test operators for adversarial attacks applied on neural text detectors.
    \item Revealing the state-of-the-art, RoBERTa-based, language analysis model is extremely vulnerable to simple mutation-based attacks.
    \item Demonstrating a random removing training strategy that improves the robustness of language analysis models significantly without requiring additional data or work.
\end{itemize}

The rest of the paper is organized as the following: We present the related works in Section~\ref{sec:related_work}. The mutation-based text generation method, adversarial attack to neural text detectors, and the RR training strategy are introduced in Section~\ref{sec:mutation_attach}. The details of the experiment and results are presented in Section~\ref{sec:experiemnts}. The paper ends with a discussion and conclusion in Section~\ref{sec:conlusion}.

\section{Related Work}
\label{sec:related_work}
\subsection{Adversarial Attacks in Machine Learning}
Adversarial attacks are a growing threat in AI and machine learning. They are also a new type of threat to cyber security, targeting the brains (i.e., machine learning models/algorithms) in the defenders (i.e., cyber security controls and protection systems) ~\cite{kaloudi2020ai,becue2021artificial}. Adversarial attacks are closely related to adversarial machine learning, a technique that attempts to fool machine learning models with deceptive data~\cite{lowd2005adversarial}. 

Adversarial attacks are classified into two categories—targeted attacks and un-targeted attacks. The former aims to fool the classifier by giving predictions for a specific target class, and the latter tries to fool the classifier by giving a wrong prediction while no specific class is targeted~\cite{qiu2019review}. 
The deceptive data are often purposely designed to cause a model to make a mistake in its predictions despite resembling a valid input to a human. Numerous ways may be used to acquire the deceptive data or adversarial samples, such as Fast Gradient Sign Attack (FGSA)~\cite{kurakin2016adversarial} and generative adversarial networks (GANs)~\cite{goodfellow2020generative,liang2019ganai}. 

In this work, we use the proposed mutation-based generation method to produce the adversarial text samples and conduct an un-targeted attack on the neural network classifiers that distinguish machine-generated text from human-written ones.

\subsection{Automatic Text Generation}
Automatic text generation is a field of study that has a long history in Natural Language Processing (NLP) that combines computational linguistics and artificial intelligence ~\cite{mann1983overview,jelinek1985markov}.
The field has progressed significantly in recent years due to the advanced neural network technologies~\cite{guo2018long,zhu2018texygen,yu2022survey}. Neural network based approaches are dominant in the field nowadays. Popular methods may include Transformer, BERT, GPT-3, RoBERTa, and their variants that are used in several text domains.

The neural network based models are often trained on large text datasets. For example, the GPT-2 model was trained on text scrapped from eight million web pages~\cite{radford2019language} and is able to generate human-like texts. Due to the high text generation performance, such methods are popular in image caption generation~\cite{vinyals2015show}, automatic text summarization~\cite{el2021automatic}, machine translation~\cite{vaswani2018tensor2tensor}, moving script-writing~\cite{zhu2020scriptwriter}, poetry composition~\cite{yi2017generating}, etc. The vast majority of automatic text generation methods focus on content generation. Though advanced control may be applied, the text is still generated in a largely open-ended fashion, where acquiring precise output is still non-trivial.

Unlike the popular neural network based approach, our method generates output in a close-ended fashion. The precise outputs are produced based on a given text sequence and a specific mutation operator. Due to the well-controlled generation process, our method can be used to systematically evaluate language analysis models.

\subsection{Neural Text Detection}
In this work, we refer to neural text detection as the detection task that distinguishes machine-generated text from human-written ones. Though neural text detection may still be under-researched compared with the imaging domain, it has attracted increasing attention over the last few years~\cite{gehrmann2019gltr,adelani2020generating,Bhatt2021detecting,solaiman2019release}. 
Various approaches have been proposed for predicting whether a text sequence has been machine-generated or not. For instance, Bhatt and Rios used linguistic accommodation in online-conversion interaction to identify whether the text was generated by a chatbot~\cite{Bhatt2021detecting}. Solaiman et al. used the probability distribution expressed by neural language models directly by computing the total probability of the text sequence of interest. If the computed probability is closer to the mean likelihood over a set of known machine-generated sequences, the text sequence is classified as machine generated~\cite{solaiman2019release}. In this study, we used the RoBERTa-based detector to demonstrate the neural text detection task. The model was based on the largest GPT-2 model (consisting of 1.5B parameters)~\cite{radford2019language}. Additionally, the model is fine-tuned to distinguish between texts being generated from the GPT-2 model and human texts. In total, 500,000 text samples were used in training~\cite{solaiman2019release,stiff2022detecting}.

\section{Mutation-Based Text Generation and Adversarial Attacks}
\label{sec:mutation_attach}

Inspired by the advances in mutation analysis in software testing, two general types of mutation operators are proposed---the character- and word-level mutation operators---for generating adversarial samples. The adversarial samples are then used to attack the state-of-the-art neural text detectors that evaluate whether the input text is created by a machine or a human.

\subsection{Mutation Operators}
\label{sec:mutation_operator}

\subsubsection{Character-Level Mutation Operators} \label{sec:cmutation}
Given a text corpus (e.g., a paragraph), $\tau$ which contains an ordered set of words, $\omega=\{\omega_1, \omega_2, ..., \omega_n\}$, and an ordered set of punctuation, $\upsilon=\{\upsilon_1, \upsilon_2, ..., \upsilon_m\}$, a mutation operator, $\mu_c(\cdot)$ is used to generate the character-level mutation of $\tau$ by replacing a given character to a close form for a specific word. Mathematically, this process is defined as:
\begin{equation}
    \omega_i' = \mu_c(\omega_i, \rho, \sigma),
    \label{eq:mse}
\end{equation}
where $\omega_i \in \omega$, $\rho$ is a letter in $\omega_i$, $\sigma$ is the mutation of $\rho$, and $\omega_i'$ is the mutation of $\omega_i$. 
After the mutation, the original $\omega$, where $\omega \in \tau$ and $\omega=\{\omega_1, \omega_2, ..., \omega_i, ..., \omega_n\}$, is changed to $\omega'=\{\omega_1, \omega_2, ..., \omega_i', ..., \omega_n\}$. The mutation text corpus is $\tau'=\{\omega',\upsilon\}$. 
For instance, given a text corpus, $\tau$, where $\tau=$ ``this is an apple". The ordered set of words is $\omega=\{$this, is, an, apple$\}$. Assume $\omega_i=$ apple, $\rho=$ a, and, $\sigma=\alpha$. Then, $\omega_i'=\alpha$pple and $\omega'=\{$this, is, an, $\alpha$pple$\}$.

\subsubsection{Word-Level Mutation Operators}
Similar to the character-level mutation (Sec~\ref{sec:cmutation}), given a text corpus (e.g., a sentence), $\tau$, a mutation operator, $\mu_w(\cdot)$, is used to generate the word-level mutation by replacing a specific word by another one. Specifically,
\begin{equation}
    \omega'' = \mu_w(\omega, \omega_j, \omega_j'),
    \label{eq:mse}
\end{equation}
where $\omega_j \in \omega$, $\omega_j'$ replaces $\omega_j$, and $\omega''=\{\omega_1, \omega_2, ..., \omega_j', ..., \omega_n\}$. 
For instance, given the same text corpus as before where: 
$\tau=$ "this is an apple". 
Assume $\omega_j=$ apple, and $\omega_j'=$ orange. 
After applying $\mu_w(\cdot)$ to $\omega=\{$this, is, an, apple$\}$, the mutation is $\omega''=\{\text{this, is, an, orange}\}$. 
Then, $\tau''=\{\omega'',\upsilon\}$.

\subsubsection{Using the Operators} 
The character- and word-level mutation operators introduced in this section are two general operators. Various specific operators may be developed based on these two general types. For instance, synonyms mutation operators can be designed using the word-level mutation operator by replacing a specific word in a text corpus with the synonym. Similarly, an adjective-removing operator can also be developed by replacing the adjectives in a text corpus with \texttt{""} (i.e., the empty string). As a proof of concept, in this work, we proposed nine sets of mutation operators in Section~\ref{sec:experiemnts}. A detailed explanation of the operators is given in Section~\ref{sec:mutation_operators}. However, users are not limited to the operators used in this study. One may easily design more operators that may be more suitable for their specific tasks under the proposed framework.

\subsection{Attacks on Neural Text Detectors}
\begin{figure}[!tb]
    \centering
    \begin{subfigure}[b]{.95\textwidth}
      \includegraphics[width=.925\textwidth]{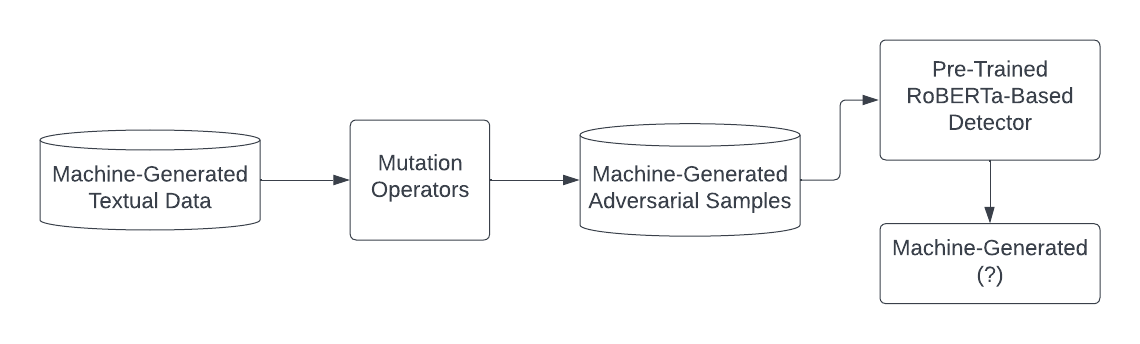}
      \caption{}
      \label{fig:attack_roberta}
    \end{subfigure}
    \begin{subfigure}[b]{.95\textwidth}
      \includegraphics[width=.925\textwidth]{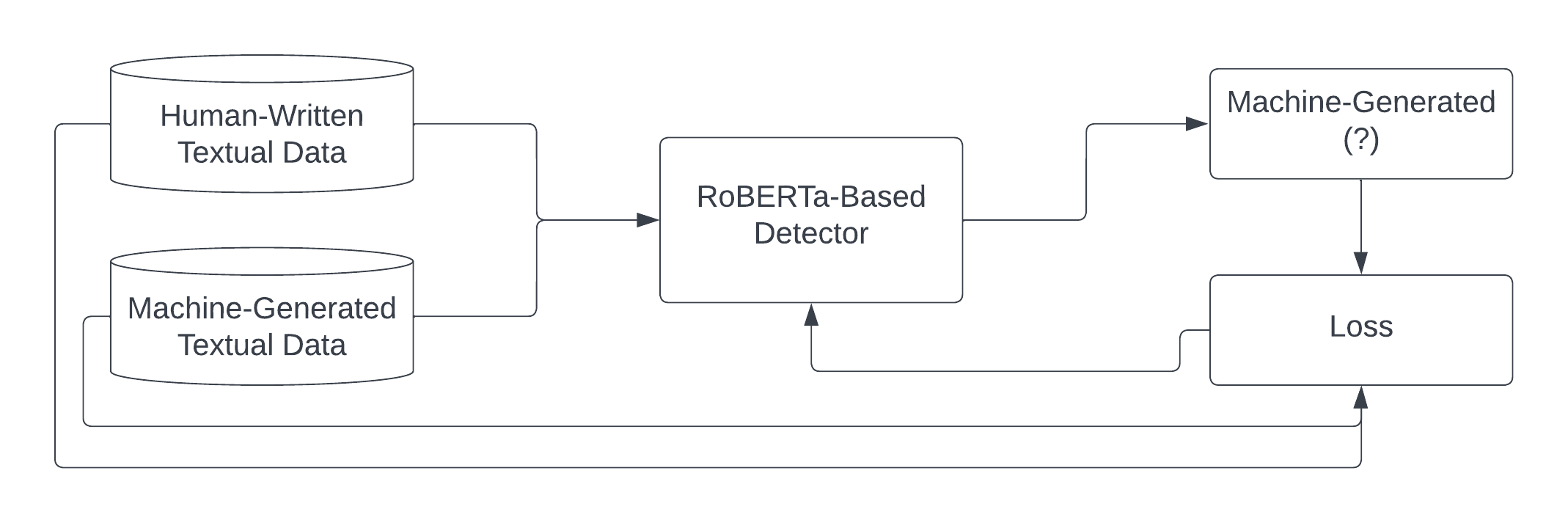}
      \caption{}
      \label{fig:roberta}
    \end{subfigure}
    \caption{(a) Adversarial attacking the pre-trained RoBERTa-based detector. (b) The RoBERTa-based detector general training procedure.}
    \label{fig:attack}
\end{figure}

Neural network language models can be trained to distinguish human-written textual content vs. machine-generated textual content. Among several existing detectors, RoBERTa-based detector~\cite{solaiman2019release} is well-known for the state-of-the-art performance on the GPT-2 generated text~\cite{stiff2022detecting,jawahar2020automatic}. 
We apply adversarial attacks to the RoBERTa-based detector using the adversarial samples generated with the mutation operators introduced in Sec~\ref{sec:mutation_operator}. The basic process is illustrated in Figure~\ref{fig:attack}.

To attack the neural text detectors, we first apply the mutation operators to a set of machine-generated textual content to generate the adversarial samples. Then, the adversarial samples are used to test a pre-trained RoBERTa-based detector (Figure~\ref{fig:attack_roberta}), which was released by OpenAI\footnote{https://openai.com}. This is accomplished by fine-tuning a RoBERTa large model with the outputs of the 1.5B-parameter GPT-2 model, following the general process summarized in Figure~\ref{fig:roberta}. 

The original RoBERTa was trained for text generation on 160 GB of text, including Books Corpus, English Wikipedia, CommonCrawl News dataset, Web text corpus, and Stories from Common Crawl~\cite{liu2019roberta}. 
To fine-tune the detector, both human-written and machine-generated textual data were fed to the detector to evaluate whether a machine-generated the text or not. 
A binary loss function was used to assess the network prediction by comparing the prediction and the ground-truth label of the input. Afterward, the loss is back-propagated to the detector for tuning the network parameters.

\subsection{Random Removing Training}
\label{sec:rr_training}
Empirically speaking, pre-trained neural network models are often less robust when handling out-of-distribution data. Research showed that relaxing model prediction by introducing a small amount of uncertainty to classification models helps increase the overall model performance~\cite{pereyra2017regularizing,muller2019when,yun2019cutmix,liang2020improved}. For instance,  Label Smoothing~\cite{pereyra2017regularizing} is a widely used technique in computer vision classification tasks. Instead of targeting label 1 for the correct class, Label Smoothing tries to predict $1-\epsilon$ for the correct class label, where $\epsilon$ is usually a small number, such as $0.01$. MixCut~\cite{yun2019cutmix} is another example, which mixes two images of two classes together into one image and predicts the mix proportionally to the number of pixels of combined images. Both Label Smoothing and MixCut demonstrated a significant performance improvement by introducing a small uncertainty to the model.

Inspired by the methods mentioned above, we propose a random removing (RR) training strategy to improve the robustness of neural text detectors by introducing a small uncertainty during the training stage (Algorithm~\ref{alg:rr}). Specifically, given a training instance, we randomly apply the word-level mutation operator $\mu_w$ where $\mu_w=(\omega, \omega_j, \texttt{""})$ to $k$ words in $\omega$. 
For each input text $\tau$, $k$ is randomly decided and $k<=floor(len(\omega)\times1/3)$, where $len(\omega)=$ indicate the number of words in the ordered list and $floor(\cdot)$ indicates the floor function. 
For instance, assume $\tau=\{\text{this is an apple}\}$. Then, $len(\omega)=4$, $floor(len(\omega)\times1/3)=1$, and $0<=k<=1$. 

\begin{algorithm}[!tb]
    \DontPrintSemicolon
    \SetAlgoLined
    \SetNoFillComment
    \caption{Random Removing Training}
    \label{alg:rr}
    
    \tcc{Require: Neural text detector model $h_\Theta(\cdot)$, training data $\mathcal{D}=\{\mathcal{T}, \mathcal{L}\}$, text corpus $\mathcal{T} = \{\tau_1, \tau_2, ..., \tau_i\}$, classification labels $\mathcal{L}=\{\iota_1, \iota_2, ..., \iota_i\}$ where $\iota \in \{0, 1\}$, loss function $\mathbb{L}(\cdot)$, word-level mutation operator $\mu_w(\cdot)$, random class $\mathbb{R}(\cdot)$, and sort function $sort(\cdot)$.}
    \;
    \For{$i \in 0$ to $len(\mathcal{T})$}{    
        $\omega, \upsilon \gets \tau_i$ \tcp*{Get the ordered word list and the punctuation list from $\tau_i$}               
        $r \gets \mathbb{R}.randInt(0,1)$ \tcp*{Get a random integer for 0 or 1.}
        \;
        \tcc{Apply mutation to $\tau_i$ if r is 1}
        \If{$r==1$} { 
            $n \gets \mathbb{R}.randInt(0,floor(len(\omega)/3)$ \tcp*{Get a random number of words} 
        \tcp*{that will be removed from $\omega$}    
            $indices \gets \{index_1, index_2, ..., index_n\}$ \tcp*{Get a list of $n$ random indices of} 
            \tcp*{words, where $0<= index_i<len(\omega)$.} 

            $sort(indices)$ \tcp*{Sort the list in an ascending order.}
            \;
            
            \tcc{Remove $n$ words}
            \For{$j \in 0$ to $n$}{
                $\omega = \mu_w(\omega, \omega_{indices[j]}, \text{""})$ \tcp*{Using the word-level mutation operator}
                \tcp*{to remove words from $\omega$}
            }
            $\tau_i \gets \{\omega, \upsilon\}$ \tcp*{Update $\tau_i$. $n$ words are removed from the original $\tau_i$}  
        }
        \;
        \tcc{Train the model}
        $logits \gets h_\Theta(\tau_i)$ \tcp*{Feed data to the model}
        $loss = \mathbb{L}(logits, \iota_i)$ \tcp*{Calculate the loss}        
        $loss.back()$ \tcp*{Back-propagation}
    }              
\end{algorithm}

\section{Experiments and Analysis}
\label{sec:experiemnts}
\subsection{Experimental Setup}
The influences of social media in our life is increasing daily. A large-scale information operation on social media may potentially lead to national crises. Automatic content generation reduces the human resource demand of launching such an operation. Thus, distinguishing machine-generated posts from those written by humans in social media posts is a critical step to defend an AI-backed information operation. With such a background in mind, we designed our experiments to simulate an oversimplified social media scenario, where social media posts may have two typical features:
\begin{itemize}
    \item The text in a post is usually relatively short. For instance, the common length of tweets is often between 25-50 characters.
    \item A good percentage of posts contain images and text. The text is often related to the image. 
\end{itemize}

\subsubsection{Experimental Dataset}
To simulate the oversimplified social media scenario mentioned above and acquire machine-generated and human-written text more easily, we used the MS COCO2017 dataset~\cite{lin2014microsoft} in our experiments. To speed up our experiments, the first 10,000 samples were selected from the dataset and used in our experiment. Each sample contains one image and five captions written by human users. We applied a pre-trained image caption generation model~\cite{shrimal2020attention} on each image in order to acquire the machine-generated text. Five captions were generated for each image. 

In total, the text dataset contains $100,000$ image captions for $10,000$ images, with 50,000 from the MS COCO2017 and 50,000 generated by us. Then, we used the original MS COCO2017 captions as human-written samples and the caption generated by us as machine-generated text. The dataset was divided into the train, val, and test sub-sets on the image-level with a ratio of 70:15:15, respectively.

\subsubsection{Models and Training Setup}
Three models are compared in this study, namely: 1) RoBERTa-Base, 2) RoBERTa-Finetune, and 3) RoBERTa-RR.
\begin{itemize}
    \item RoBERTa-Base: The RoBERTa-based detector that was originally released by OpenAI. 
    
    \item RoBERTa-Finetune: A finetuned RoBERTa-based detector using our training set. All the embedding layers were frozen during the training. We only optimized the classifier as part of the model. 
    
    \item RoBERTa-RR: Another finetuned RoBERTa-based detector that followed the same setup of the RoBERTa-Finetune model but trained using the RR training strategy.
\end{itemize}

All the experiments in this paper were conducted on an NVIDIA T100 GPU card. Our test set was used to evaluate the performance of all three models. The HuggingFace (version 2.9.1) implementation of \texttt{RobertaForSequenceClassification} with \texttt{roberta-large} weights~\cite{wolf2019transformers} was used as the architecture of the RoBERTa-based detector. The OpenIA pre-trained weights\footnote{https://openaipublic.azureedge.net/gpt-2/detector-models/v1/detector-large.pt} were used for RoBERTa-Base and used to initialize the RoBERTa-Finetune and RoBERTa-RR models. Both finetune models were trained for 50 epochs using our train set. The best checkpoint was selected based on the evaluation performed on the val set. The test set was applied for the final evaluation. We padded each input to have a maximum length of 50. The AdamW~\cite{loshchilov2017decoupled} optimizer with a learning rate of $1e^{-4}$, a batch size of 512, and the cross-entropy loss was used during the training.

\subsubsection{Mutation-based Adversarial Attacks}
\label{sec:mutation_operators}

We developed nine sets of mutation operators using the character- and word-level operator introduced in Section~\ref{sec:mutation_operator}, including six sets of character-level operators and three sets of word-level operators. The output text from the nine sets of operators was then used as adversarial samples to attack the three models.

The character-level operators can be categorized into two groups: 1) $\mu_c(\omega_i, \text{a}, \alpha)$ and 2) $\mu_c(\omega_i, \text{e}, \epsilon)$. 
Each group contains three sets of mutation operators, the mutation operator of articles, the mutation operator of adjectives, and the mutation operator of adverbs. In total, six sets of character-level operators were formed. The operators of articles were focusing on the three articles---a, an, the. The operators of adjectives were focusing on 527 common adjectives provided by Missouri Baptist University (MBU) Academic Success Center\footnote{https://www.mobap.edu/wp-content/uploads/2013/01/list\_of\_adjectives.pdf}. The operators of adverbs were focusing on 255 common adverbs also provided by MBU\footnote{https://www.mobap.edu/wp-content/uploads/2013/01/list\_of\_adverbs.pdf}. 

The word-level mutation operators were applied to remove certain words, $\mu_w(\omega, \omega_i, \texttt{""[empty string]})$. Following the setup of the character-level mutation operators, the $\omega_i$ used in $\mu_w$ were also selected from the three types of words---articles, adjectives, and adverbs. The same word lists were used to build the word-level operators.

\begin{table}[!tb]
    \setlength{\tabcolsep}{2.2pt}
	\centering
	\caption{Detailed performance of RoBERTa-Base, RoBERTa-Finetune, and RoBERTa-RR on different classification cases.}
	\begin{tabular}{c||c||c|c|c|c|c|c|c} \hline \hline
	\textbf{Metric} & \textbf{Model} & \textbf{H vs M} & \textbf{H vs M\textsubscript{mwr}} & \textbf{H vs M\textsubscript{mwj}} & \textbf{H vs M\textsubscript{mwd}} & \textbf{H vs M\textsubscript{mcr}} & \textbf{H vs M\textsubscript{mcj}} & \textbf{H vs M\textsubscript{mcd}} \\\hline    

    \multirow{3}{*}{\textbf{AUC}} & \textbf{RoBERTa}  & $0.6381$ & $0.2488$ & $0.3695$ & $0.2591$ & $0.0676$ & $0.0676$ & $0.0714$ \\\cline{2-9}
    & \textbf{Finetune} & $\bf{0.8626}$ & $0.4774$ & $ 0.4825$ & $0.4619$ & $0.3237$ & $0.3237$ & $0.1958$ \\\cline{2-9}
    & \textbf{RR} & $0.8520$ & $\bf{0.6496}$ & $\bf{0.6486}$ & $\bf{0.6655}$ & $\bf{0.3617}$ & $\bf{0.3617}$ & $\bf{0.2955}$\\\hline \hline	

    \multirow{3}{*}{\textbf{ACC}} & \textbf{RoBERTa}  & $\bf{0.5892}$ & $0.3504$ & $0.4460$ & $0.3892$ & $0.3338$ & $0.4213$ & $0.3686$ \\\cline{2-9}
	& \textbf{Finetune} & $0.5616$ & $0.4948$ & $0.5923$ & $0.5350$ & $0.4921$ & $0.5875$ & $0.5334$ \\\cline{2-9}
    & \textbf{RR} & $0.5625$ & $\bf{0.4999}$ & $\bf{0.5989}$ & $\bf{0.5433}$ & $\bf{0.4930}$ & $\bf{0.5886}$ & $\bf{0.5344}$\\\hline \hline

    \multirow{3}{*}{\textbf{F1}} & \textbf{RoBERTa} & $0.6169$ & $0.5046$ & $0.5877$ & $0.5400$ & $0.4982$ & $0.5771$ & $0.5318$\\\cline{2-9}
	& \textbf{Finetune} & $0.6918$ & $0.6608$ & $0.7424$ & $0.6964$ & $0.6596$ & $0.7401$ & $0.6957$ \\\cline{2-9}
    & \textbf{RR} & $\bf{0.6926}$ & $\bf{0.6635}$ & $\bf{0.7459}$ & $\bf{0.7006}$ & $\bf{0.6604}$ & $\bf{0.7410}$ & $\bf{0.6966}$ \\ \hline\hline

	\end{tabular}
	\label{table:accuracy}
\end{table}

\subsection{Results and Analysis}
\begin{figure}[!tb]
    \centering
    \begin{subfigure}[b]{.475\textwidth}
      \includegraphics[width=1\textwidth]{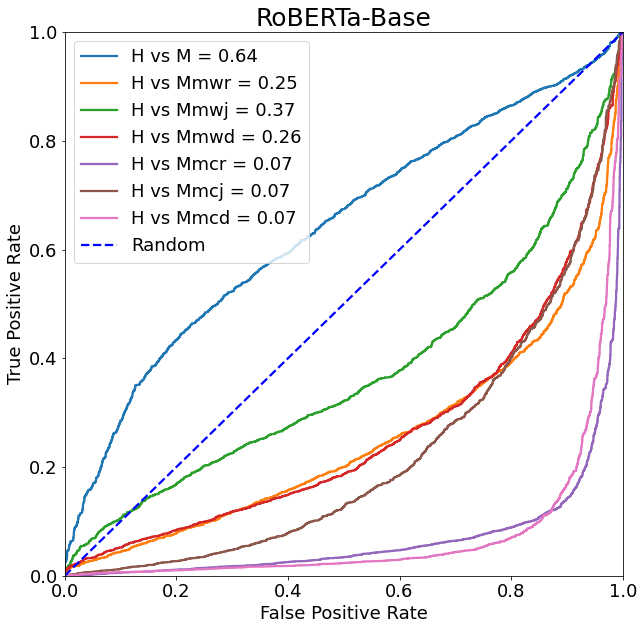}
      \caption{}
      \label{fig:result_roberta}
    \end{subfigure}~~~~~~~~~
    \begin{subfigure}[b]{.475\textwidth}
      \includegraphics[width=1\textwidth]{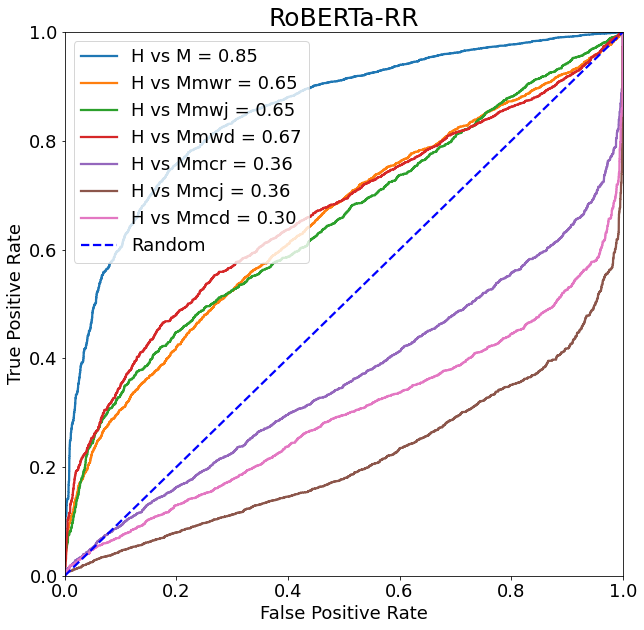}
      \caption{}
      \label{fig:result_rr}
    \end{subfigure}
    \caption{The area under the receiver operating characteristic (AUC) curve for RoBERTa-Base and RoBERTa-RR on different classification cases.}
    \label{fig:result_auc}
\end{figure}

\begin{figure}[!tb]
    \centering
    \includegraphics[width=1\textwidth]{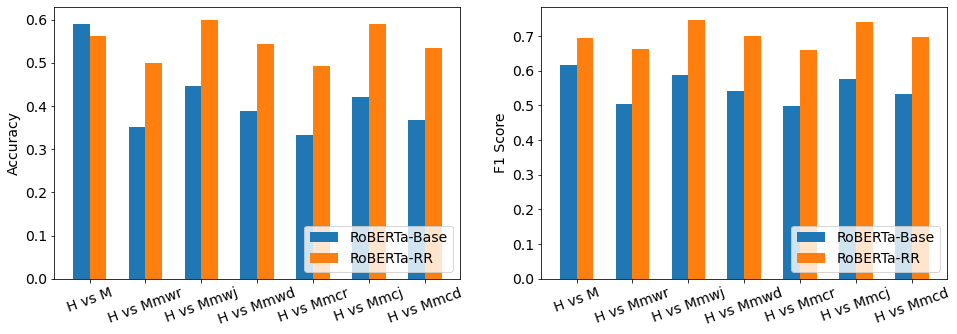}
    \caption{ The accuracy (left) and F1 score (right) for RoBERTa-Base and RoBERTa-RR on different classification cases.}
    \label{fig:result_f1_acc}
\end{figure}

Table~\ref{table:accuracy} shows the detailed performance of the three models over seven binary classification tasks. The performance of RoBERTa-Base and RoBERTa-RR is also summarized in Figures~\ref{fig:result_auc} and~\ref{fig:result_f1_acc}. The results showed that the RoBERTa-Base performed poorly on separate machine-generated text from the human-written ones, with the performance not much better than a random guess. After applying mutation operators to the machine-generated text, the performance is even worse. The result also shows that finetuning the RoBERTa-Base using our training set improves the performance significantly. The proposed RR training strategies can further improve the performance for almost all the classification tasks and metrics. The rest of this section gives a detailed explanation of the experimental tasks and the results.

In total,  seven binary classification tasks are evaluated in this study, namely Human-Written text vs. Machine-Generated text and Human-Written text vs. six sets of Machine-Generated text mutations, separately. In the table and figures, we use \textit{H} to indicate Human-Written Text, \textit{M} to indicate Machine-Generated Text, and subscript \textit{mxy} to indicate mutations and their specific operators. The second letter in the subscript indicates the level of the mutation that can be either word-level (denoted as \textit{w}) or character-level (denoted as \textit{c}). The third letter in the subscript indicates the types of words that the mutation operator is applied on, such as \textit{r} for articles, \textit{j} for adjectives, and \textit{d} for adverbs. For instance, \textit{H vs M} means Human-Written text vs. Machine-Generated text, \textit{H vs M\textsubscript{mwr}} means Human-Written text vs. Machine-Generated text Mutations w/ the word-level mutation operators being applied to the articles, and \textit{H vs M\textsubscript{mcd}} means Human-Written text vs. Machine-Generated text Mutations w/ the character-level mutation operators being applied to the adjectives. 

We evaluate the performance of each model for each classification task using the area under the receiver operating characteristic curve (AUC), the accuracy (ACC), and the F1 score (F1). 
Table~\ref{table:accuracy} shows the RoBERTa-Base model has a about $59\%$ accuracy and about $0.62$ F1 score for Human vs Machine, which is not much better than a random guessing. The result is also within the range of a previous study~\cite{stiff2022detecting}. When applying the mutation operators to the machine generated text, RoBERTa-Base gets an even worse performance with an average of $0.1583$ on AUC, $38.47\%$ on accuracy, and $0.4589$ on F1 Score.

After finetuning the model using our training set, RoBERTa-Finetune boosts the performance of all the metrics that involve mutation operators. Note that no mutation operators are applied to the training set. On average, the RoBERTa-Finetune is able to improve the performance on the mutation test to $0.4544$ on AUC, $53.92\%$ on the accuracy, and $0.6992$ on F1, which is $187\%$, $40.16\%$, and $52.36\%$ improvement on AUC, ACC, and F1, respectively. Furthermore, when the RR strategy is used in finetuning, the performance involving mutation operators are further improved by an average of $9.40\%$ on AUC to 0.4971, $0.7\%$ on ACC to $54.30\%$, and $0.3\%$ on F1 to 0.7013.


\section{Discussions, Conclusions and Future Directions}
\label{sec:conlusion}
Due to the progress of advanced automatic content generations, language models may produce human-like text that benefits society, from question-answering to AI-driven education. However, there is also a risk that such advanced intelligent techniques may be used by malicious actors for nefarious activities at large-scale, from spreading misinformation to targeting specific groups (e.g., for political agenda). A robust detector that can separate machine-generated text from human-written ones is the first step to defend against such newly emerged AI-powered cyber threats. Unfortunately, how researchers may systematically evaluate the robustness of a neural text detector is still being determined in the literature. 
Thus, we propose a mutation-based text generation framework that produces textual output in a well-controlled and close-ended environment. The precise output of our approach provides a novel way to systematically evaluate the robustness of language analysis models. 

In this study, we use the outputs of our mutation-based text generation method as adversarial samples to attack the state-of-the-art RoBERTa-based detector for separating human-written text from machine-generated ones. The result shows that the detector has significant flaws, which are extremely vulnerable to simple adversarial attacks, such as replacing the English letter ``a" with the Greek letter ``$\alpha$". 
To improve the robustness of the detector, we proposed a random removing (RR) training strategy that introduces uncertainty at the finetuning stage, which significantly improves the model' robustness on all nine types of attacks (up to $33.74\%$ on AUC). 
However, we also believe this issue should be better addressed at the feature level because the text-level changes cause changes in the tokenization stage, leading to a different embedding vector.
The contextual embedding characteristics of RoBERTa may make this vector change more dramatic. As an advanced technique, contextual embedding helps better understand the content by generating different embeddings of the same word base on the context. Thus, when feeding the adversarial samples into the RoBERTa-based detector, contextual embedding layers might also produce different embeddings of the non-changed words. In this case, the distance between the original and adversarial samples may increase even more in the feature spaces. Thus, one future direction of this work is reducing the feature space differences between the original and adversarial samples.

One thing worth noting is that the proposed method is a general-purpose text-generation method, which is not limited to any specific application domain. The output of our framework can be used in evaluating the robustness of any downstream applications that take text sequences as input, such as ChatBot detection, SQL injection, and software debugging. 
In addition, researchers may also design their own mutation operators that better fit their specific tasks under our framework. 

In conclusion, we proposed a mutation-based text generation framework that produces close-ended, precise text. The output of our framework can be used to evaluate language analysis models that take text sequences as input. 
We demonstrate the framework for adversarial attacking to the RoBERTa-based neural text detector. The results not only show the detector is extremely vulnerable to simple adversarial attacks but also lead to an insightful analysis of the flaws. We believe the proposed framework will be useful for those seeking systematic and insightful analysis of language models.

\bibliographystyle{elsarticle-num}
\bibliography{bibfile}







\end{document}